\documentclass[11pt,a4paper]{article}
\usepackage[utf8]{inputenc}
\usepackage[english]{babel}
\usepackage[left=1.8cm,right=1.8cm,top=2.cm,bottom=2.cm]{geometry}

\usepackage{amsmath}
\usepackage{amsfonts}
\usepackage{dsfont}
\usepackage{amssymb}
\usepackage{graphicx}
\usepackage{epstopdf}
\usepackage{subfigure}
\usepackage{hyperref}
\usepackage{multirow}
\usepackage[ruled,vlined]{algorithm2e}
\usepackage{setspace}
\usepackage{amsthm}
\usepackage{color}
\usepackage{authblk}
\usepackage{subfigure}

\usepackage{verbatim} 

\newtheorem{lem}{Lemma}

\theoremstyle{definition}
\newtheorem{definition}{Definition}

\theoremstyle{remark}
\newtheorem{remark}{Remark}

\newcommand{\E}{\ensuremath{\mathbb{E}}}
\newcommand{\Z}{\ensuremath{\mathbb{Z}}}
\newcommand{\Q}{\ensuremath{\mathbb{Q}}}
\newcommand{\R}{\ensuremath{\mathbb{R}}}

\newcommand{\Expect}{\ensuremath{{\rm I\kern-.3em E}}}
\newcommand{\Var}{\ensuremath{\mathrm{Var}}}

\newcommand{\chf}{\hat{f}}
\newcommand{\sinc}{\text{sinc}}
\renewcommand{\d}{{\rm d}}      

\title{SWIFT calibration of the Heston model}
\date{}

\author[]{Eudald Romo$^1$ and Luis Ortiz-Gracia$^{2,}$\footnote{Corresponding author: luis.ortiz-gracia@ub.edu} \\ $^1$Trading Department, Xanadu Trading Limited, Barcelona, Spain  \\ $^2$Department of Econometrics, Statistics and Applied Economics, University of Barcelona, Barcelona, Spain}

\begin{document}

\maketitle

\begin{abstract}
In the present work, the European option pricing SWIFT method is extended for Heston model calibration. The computation of the option price gradient is simplified thanks to the knowledge of the characteristic function in closed form. The proposed calibration machinery appears to be extremely fast, in particular for a single expiry and multiples strikes, outperforming the state-of-the-art method we compare with. Further, the a priori knowledge of SWIFT parameters makes possible a reliable and practical implementation of the presented calibration method. A wide set of stress, speed and convergence numerical experiments is carried out, with deep in-the-money, at-the-money and deep out-of-the-money options for very short and very long maturities.
\end{abstract}

 \vspace{0.5cm}



\section{Introduction}
The Heston model is a well-known stochastic volatility (SV) model for driving the dynamics of the assets. In order to use the Heston model, we need to calibrate its five parameters to real-market data. 
The goal of calibrating a model using market data is to estimate the model parameters in such a way that, when it is used for option valuation with an appropriate option valuation method, it yields prices similar to the real market ones. Calibration is an important task that requires efficient numerical methods. It encompasses a machinery for option pricing as well as an optimization procedure, aiming at minimizing the differences between the market option prices and the prices given by the valuation method. In regard to the pricing, we use the SWIFT method presented in \cite{Ortiz-Gracia2016} for European options. It belongs to the class of Fourier inversion methods and has already been used for pricing Bermudan, barrier and Asian options (see \cite{mar17,leitao18}). The power of SWIFT method partly relies on the knowledge of the characteristic function (ChF) in analytical form. Since the ChF associated to the Heston model is known in closed form, we can tackle the optimization problem by means of the gradient based Levenberg-Marquardt (LM) algorithm \cite{mor78}. For sake of comparison, we consider the state-of-the-art calibraton method of \cite{cui17}, which is based on the Fourier inversion pricing method of \cite{Heston1993} and it also uses the LM optimization algorithm. The main contributions of this work are the following.
\begin{itemize}
\item We extend the SWIFT method to the calibration problem by deriving the option price gradient.
\item We implement and test the speeding up techniques mentioned in \cite{Ortiz-Gracia2016} based on multiple strikes valuation.
\item We propose a novel method for calibrating the Heston model with a set of options with certain fixed strikes that can be used later on for arbitrary strikes by interpolation.
\item We develop and implement speeding up techniques for the option price gradient.
\end{itemize}
We carry out a wide variety of stress, speed and convergence tests with at-the-money (ATM), deep in-the-money (ITM) and deep out-the-money (OTM) options, ranging from very short to very long maturities. The results show that SWIFT is extremely fast for sets of options with a single expiry and different strikes, being about ten times faster thant the calibration method of \cite{cui17}. For options with a fixed number of maturities and a fixed number of strikes per maturity, both methods perform similar in terms of speed, beeing more robust the SWIFT method, thanks to the possibility of a priori selecting the parameters of the pricing machinery. This last feature, makes SWIFT method a reliable and practical methodology for real-time updating of Heston model calibration.

The paper is organized as follows. We define the calibration and the valuation problem in Section 1. The Heston model, the valuation method of \cite{cui17} and the calibration challenges are presented in Section 2. Section 3 is devoted to the SWIFT method and its speeding up features. In Section 4, we put forward the calibraton problem with all the mathematical details, and we test our proposed method through the numerical experiments of Section 5. Section 6 concludes and gives some pointers for future research.

\subsection{Option valuation}\label{chapter:option_valuation}
The option valuation will be tackled under the framework of the expected discounted payoff pricing formula that we next recall.
Consider a European option contract with strike price $K$, expiring at time $T$, with $\tau := T - t$ the time to maturity, and $S_{t}$ the price of its underlying asset at time $t$. Then, if one considers state variables $x$ and $y$ that fully describe $S_t$ and the random variable $S_T$ respectively, the general pricing formula becomes,
\begin{equation}
\label{eq:integral_option_valuation}
v(x,t)  = e^{-r \tau}\E^{\Q}(v(y, T)|x)  = e^{-r\tau} \int_{\R}v(y,T)f(y|x)dy,
\end{equation}
where $v(x,t)$ denotes the option value at time $t$, $v(y,T)$ is the payoff, $r$ the risk-neutral interest rate, $\E^{\Q}$ the expectation under the risk-neutral measure, and $f(y|x)$ is the probability density of $y$ given $x$. A common choice for the state variables $x$ and $y$ is to define,
\begin{align}
\label{eq:y}
y = \ln \left(\dfrac{S_T}{K} \right), \\
\label{eq:x}
x = \ln \left(\dfrac{S_t}{K} \right).
\end{align}
More precisely, given the states variables choice, we can express the payoff of a European option in log-asset price as,
\begin{equation}\label{eq:european_payoff}
v(y,T)=\left[ \alpha \cdot K \left( e^y-1 \right) \right]^+:=K \cdot g(y), \quad \text{with,} \quad 
\alpha=\begin{cases} 1, & \text{for a call}, \\ -1, & \text{for a put}, \end{cases}
\end{equation}
where $g(y):=\max \left(\alpha \left(e^y-1\right),0 \right) $ denotes the strike-free payoff.



\section{Heston dynamics and calibration issues} \label{chap:heston_model}

The widely used Black and Scholes (BS) model fails to capture essential well-known properties of the real world market dynamics of the underlying return distributions, as its high kurtosis, its negative skew, the correlation between the underlying price and its volatility, the risk premium investors give deep (ITM) or (OTM) options, etc. All these properties result in the well-known BS implied volatility surface. SV models, treat both the underlying price and its volatility as (potentially correlated) stochastic processes, which allows to better capture some of these properties. One of the first and most well-known SV models is the Heston model \cite{Heston1993}, defined by the following system of stochastic differential equations.

\begin{definition} The Heston model price-volatility dynamics is defined by,
\begin{align*}
& dS_t = \mu S_t dt + \sqrt{\nu_t} S_t dW_t^{(1)}, \\& d\nu_t = \kappa(\overline{\nu} - \nu_t) dt + \sigma \sqrt{\nu_t}dW_t^{(2)},
\end{align*}
where $W_t^{(1)}$ and $W_t^{(2)}$ are two correlated Wiener processes $dW_t^{(1)}dW_t^{(2)} = \rho dt$ and $\nu_t$ is the variance of the underlying asset price at time t. Then, if one specifies the initial value of the variance $\nu_0$, the model is properly defined. From now on, $\boldsymbol{\theta} := (\nu_0, \overline{\nu}, \rho, \kappa, \sigma)^T$ will refer to the vector of model parameters.
\end{definition}
Several works have shown the relationship between the Heston model parameters and the shape of the implied volatility surface \cite{cla11, gat06, gil12, jan11} necessary to obtain the same prices with a BS model. It can be summarized as follows,

\begin{itemize} 
\item $\nu_0$ controls the position of the volatility surface,
\item $\rho$ controls its skewness,
\item $\kappa$ and $\sigma$ control the convexity of the surface,
\item $\kappa(\nu_0 - \overline{\nu})$ controls the term structure of implied volatility.
\end{itemize}

A method for calibrating the Heston model is presented in \cite{cui17}. This method starts from the expression of the price of a European call option presented in the original work by Heston \cite{Heston1993}, and it is adapted here in terms of the state variables x and y defined in Section \ref{chapter:option_valuation},

\begin{equation} 
\label{eq:option_valuation_final_form}
v(x, \tau) = K e^x P_1(\boldsymbol{\theta}; x, \tau) - K e^{-r \tau} P_2(\boldsymbol{\theta}; x, \tau),
\end{equation}
where $P_1$ and $P_2$ are defined as,
\begin{align}\label{eq:P1}
&P_{1}(\boldsymbol{\theta} ; x, \tau)=\frac{1}{2}+\frac{1}{\pi} \int_{0}^{\infty} \operatorname{Re}\left(\frac{e^{i u x}}{i u} \hat{f}(-u+i)\right) \mathrm{d}u, \\
\label{eq:P2}
&P_{2}(\boldsymbol{\theta} ; x, \tau)=\frac{1}{2}+\frac{1}{\pi} \int_{0}^{\infty} \operatorname{Re}\left(\frac{e^{iux}}{i u} \hat{f}(-u)\right) \mathrm{d}u,
\end{align}
and $\hat{f}(u)$ is the initial state independent ChF of the process.
It is worth remarking that, as we will see in Section \ref{sec:swift}, the SWIFT method will benefit from the fact that the ChF $\hat{f}(u)$ does not depend on the initial state variable.

\begin{remark}
The dependence of the ChF in time and the model parameters is omitted for readability.
\end{remark}
\begin{remark}
The expression (\ref{eq:option_valuation_final_form}) omits the dividend yield term $q$, which appears in \cite{cui17} but is assumed to be 0 here for readability. The results presented here are valid for any constant value $q$.
\end{remark}
\begin{remark}
Typically, the ChF of a random variable with density function $f$ is defined as $\tilde{f}(u)=\int_{\R} f(x) e^{iux} dx$. However, to be consistent with \cite{Ortiz-Gracia2016}, it is defined in this work as $\hat{f}(u)=\int_{\R} f(x) e^{-iux} dx$. We can see that there is a sign difference in all the $u$-dependent equations and expressions in \cite{cui17}.
\end{remark}
We write expression (\ref{eq:option_valuation_final_form}) in a more compact form in the following lemma,

\begin{lem} [Heston's pricing method]\label{lem:Hestonpricing} Let $V(\theta ; x, \tau)$ be the price of a European call option with strike $K$ and Heston dynamics, given by expression (\ref{eq:option_valuation_final_form}). If we define $\hat{f}(u; x) = e^{-iu x}\hat{f}(u)$, then,
\begin{equation} \label{eq:heston_analytic}
V(\theta ; x, \tau)= K \left[ \frac{1}{2}\left(e^x- e^{-r \tau}\right) 
 + \frac{e^{-r \tau}}{\pi}\int_{0}^{\infty} \operatorname{Re}\left( \frac{\hat{f}(-u+i; x) - \hat{f}(u; x)}{i u}\right) du\right].
\end{equation}
\begin{proof}
Having into account that  $S_t/K = e^{\log(S_t/K)} = e^x$ then, from expression (\ref{eq:option_valuation_final_form}), (\ref{eq:P1}) and (\ref{eq:P2}) we have,
\begin{equation*}
\begin{aligned}
V(\theta ; x, \tau)= K \frac{1}{2}\left(e^x- e^{-r \tau}\right) + K\frac{e^{-r \tau}}{\pi}& \left[\int_{0}^{\infty} \operatorname{Re}\left(\frac{e^{i (u - i) x}}{i u} \hat{f}(-u+i)\right) \mathrm{d} u\right. 
\left.- \int_{0}^{\infty} \operatorname{Re}\left(\frac{e^{i u x}}{i u} \hat{f}(u)\right) \mathrm{d} u\right].
\end{aligned}
\end{equation*}
Finally, since $\hat{f}(u; x) = e^{-iu x}\hat{f}(u)$, then the result follows.
\end{proof}
\end{lem}

\subsection{Calibration challenges} \label{subsec:calibration_difficulties}
As opposed to simpler 1-dimensional models, Heston model calibration is a multidimensional optimization problem with 5 degrees of freedom given by $\boldsymbol{\theta} := (\nu_0, \overline{\nu}, \rho, \kappa, \sigma)^T$. Furthermore, the structure of this optimization problem is not known.
According to \cite{cui17} no consensus exists among researchers regarding whether the objective function of this optimization problem is convex or not. Some results point to a non-convex function, as the calibration methods proposed in \cite{che07, mik03} (which yielded different results for different initial points) and one must use long or short term approximations and rules to provide a convenient initial guess. Recent research \cite{ge12} claims to provide methods that reach a unique solution independently of the initial point which, according to that study, indicates some structure that, even if not necessarily convex, tends to lead an initial guess to a stationary result.
There is also no consensus on whether the problem always has a single  optimum. In particular, it is known that there exist dependencies between the parameters that yield similar results. For example, $\lim_{t \rightarrow \infty} \Var(\nu_t) = \dfrac{\sigma^2 \overline{\nu}}{\kappa}$ (where $\Var(\cdot)$ refers to the variance operator), so large values of $\kappa$ and $\sigma$ can provide a model that prices options similarly to one with proportionally smaller values of these two parameters. The work by \cite{cui17} claims that this results in the objective function of the optimization problem being flat close to the optimum.

As said above, there is no guarantee that a gradient-based method converges to the global optimum of the model parameters, but even obtaining a local optimum has been traditionally difficult. Many papers in the literature uses numerical gradients (see \cite{ge12}) for these methods when trying to solve the Heston calibration problem (which are less accurate and more computationally consuming), because no simple analytical gradients were available and the ones obtained with symbolic algebra packages from the expressions of the ChF were intractable. Prior to \cite{cui17}, the existing methods could be summarized as follows.

\begin{itemize}
\item{\textbf{Heuristic based models.}}
Using the relationships outlined above, some works reduce the dimension of the optimization problem by assuming some values or relationships between the parameters from the observation of a specific volatility surface. For example \cite{gat06} sets $\nu_0$ to the short-term ATM implied variance obtained by using a BS model, a heuristic further justified by \cite{che07}, where the linearity between $\nu_0$ and the BS implied volatility was verified for short maturities (less than 2 months). Other heuristics used in the industry are $\kappa = \dfrac{2.75}{\tau}$ and setting $\overline{\nu}$ to the BS short term volatility \cite{cla11}. These assumptions may restrict the optimization problem domain and exclude the optimum.
\item{\textbf{Stochastic methods.}}
They are typically used in combination with deterministic search methods, as the Nelder-Mead simplex method \cite{lag98} and would avoid the pitfalls of the gradient-based methods if the optimization problem is not convex. Some examples are used in \cite{che07} and differential evolution and particle swar are used in \cite{gil12_2}.
These methods are too computationally expensive for real-time use as \cite{fer13} that employs GPU computations to calibrate options using a SV model called SABR, and it took 421.72 seconds to calibrate 12 instruments with a tolerance of $10^{-2}$ using 2 NVIDIA Geforce GTX470 GPUs.
\end{itemize}
In this work, we consider the analytical expression for the ChF provided in \cite{cui17}.
\subsection{The characteristic function}

For long-term maturities, \cite{kal06} shows that the original ChF provided in \cite{Heston1993} has discontinuities as $u$ increases, which can lead to numerical problems. The source of these discontinuities was discussed in \cite{Albrecher2007}, and an alternative expression which was continuous in the full parameters space was presented in \cite{sch04}.
A more compact version of the ChF was later derived in \cite{rol10} from the moment generating function of the process. This expression also has the benefit of having simpler analytical expressions of the gradient of the ChF than in previous expressions, but it also presents discontinuities as $u$ increases. Finally, an expression with both, simple derivatives and continuity in the full parameters domain, was provided in \cite{cui17},

\begin{equation} \label{eq:char_cui}
\hat{f}(u) = \exp \left(-i u r \tau + \dfrac{\kappa \bar{\nu} \rho \tau i u}{\sigma}-A + \dfrac{2 \kappa \bar{\nu}}{\sigma^{2}}D \right),
\end{equation} where,
\begin{equation}\begin{array}{l}
\xi:=\kappa + \sigma \rho i u, \\
d:=\sqrt{\xi^{2}+\sigma^{2}\left(u^{2} - i u\right)}, \\
A_{1}:=\left(u^{2} - i u\right) \sinh \frac{d \tau}{2}, \\
A_{2}:=\frac{d}{\nu_{0}} \cosh \frac{d \tau}{2}+\frac{\xi}{\nu_{0}} \sinh \frac{d \tau}{2}, \\
A:=\frac{A_{1}}{A_{2}}, \\
B:=\frac{d e^{\kappa \tau / 2}}{\nu_{0} A_{2}}, \\
D := \log \frac{d}{\nu_{0}}+\frac{(\kappa-d) \tau}{2}-\log \left(\frac{d+\xi}{2 \nu_{0}}+\frac{d-\xi}{2 \nu_{0}} e^{-d \tau}\right).
\end{array}\end{equation}
Further, its gradient with respect to the Heston model parameters $\boldsymbol{\theta} = (\nu_0, \overline{\nu}, \sigma, \kappa, \rho)^{T}$ is given by,
\begin{equation}
\nabla \hat{f}(u) = \boldsymbol{h}(u) \hat{f}(u),
\end{equation}
with $\boldsymbol{h}(u) = [h_{\nu_0}(u), h_{\overline{\nu}}(u), h_\sigma(u), h_\kappa(u), h_\rho(u)]^T$ and,
\begin{align*}
h_{\nu_0}(u)& =-\frac{A}{\nu_{0}}, \\
h_{\overline{\nu}}(u)& =\frac{2 \kappa}{\sigma^{2}} D+\frac{\kappa \rho \tau i u}{\sigma}, \\
h_{\sigma}(u)& =-\frac{\partial A}{\partial \rho}+\frac{2 \kappa \bar{\nu}}{\sigma^{2} d}\left(\frac{\partial d}{\partial \rho}-\frac{d}{A_{2}} \frac{\partial A_{2}}{\partial \rho}\right)+\frac{\kappa \bar{\nu} \tau i u}{\sigma}, \\
h_{\kappa}(u)& =-\frac{1}{\sigma i u} \frac{\partial A}{\partial \rho}+\frac{2 \bar{\nu}}{\sigma^{2}} D+\frac{2 \kappa \bar{\nu}}{\sigma^{2} B} \frac{\partial B}{\partial \kappa}+\frac{\bar{\nu} \rho \tau i u}{\sigma}, \\
h_{\rho}(u)& =-\frac{\partial A}{\partial \sigma}-\frac{4 \kappa \bar{\nu}}{\sigma^{3}} D+\frac{2 \kappa \bar{\nu}}{\sigma^{2} d}\left(\frac{\partial d}{\partial \sigma}-\frac{d}{A_{2}} \frac{\partial A_{2}}{\partial \sigma}\right)-\frac{\kappa \bar{\nu} \rho \tau i u}{\sigma^{2}},
\end{align*}
where the partial derivatives of A, $A_2$, B, and d are given in \cite{cui17} and can be seen in Appendix \ref{app:cui_extra}.

\section{European option valuation and calibration with SWIFT}\label{sec:swift}
In this section we give a brief overview on the SWIFT method, originally developed for pricing European options in \cite{Ortiz-Gracia2016}. In this work the method will be extended to European options calibration. For sake of completeness a section is devoted to the basic theory on Shannon wavelets.

\subsection{Multi-resolution analysis and Shannon wavelets}
Consider the space of square-integrable functions, denoted by $L^{2}(\R)$, where,
\begin{equation*}
L^{2}(\R) = \left\lbrace f: \int_{-\infty}^{+\infty} \left| f(x)\right|^2 dx < \infty \right\rbrace.
\end{equation*}
Then we can define a useful structure for function approximation called \emph{multi-resolution analysis} (MRA). Let us start with a family of closed nested subspaces in $L^{2}(\R)$,
\begin{equation*}
				\ldots \subset \mathcal{V}_{-2} \subset \mathcal{V}_{-1} \subset \mathcal{V}_{0} \subset \mathcal{V}_{1} \subset \mathcal{V}_{2} \subset \ldots,
\end{equation*}
			where,
\begin{equation*}
				\displaystyle\bigcap_{m\in\mathbb{Z}}{\mathcal{V}_{m}}=\{0\} \ , \hspace{1cm} \displaystyle\overline{\bigcup_{m\in\mathbb{Z}}{\mathcal{V}_{m}}}=L^{2}(\R),
\end{equation*}
			and,
\begin{equation*}
				f(x) \in \mathcal{V}_{m} \Longleftrightarrow f(2x) \in \mathcal{V}_{m+1}.
\end{equation*}	
If these conditions are met, then there exists a function $\phi \in \mathcal{V}_{0}$, known as \emph{scaling function} that generates a family of orthonormal bases of $\mathcal{V}_{m}$, denoted $\{\phi_{m,k}\}_{k\in\mathbb{Z}}$,
\begin{equation*}
\phi_{m,k}(x)=2^{m/2}\phi(2^{m}x-k).
\end{equation*}
These families allow to approximate any $f \in L^{2}(\R)$ with increasing resolution by means of the projection map $\mathcal{P}_{m}:L^{2}(\R) \rightarrow \mathcal{V}_{m}$,
\begin{equation*}
P_m f(x) = \sum_{k \in \Z} D_{m,k} \phi_{m, k}(x),
\end{equation*}
where $D_{m,k} = \left\langle f,\phi_{m, k}\right\rangle$, $\left\langle f,g\right\rangle = \int_\R f(x) \overline{g(x)} dx$, and $\overline{\cdot}$ is the complex conjugate operator.

Increasing the considered number of elements of the finite family will increase the resolution of the approximation, converging to a perfect representation when all the functions of the original family are used (see \cite{tour}). Apart from increasing the resolution by means of $m$, wavelets can be moved by means of $k$ and stretched or compressed by means of $m$ to represent local properties of a function. A basic reference on wavelets is \cite{dau92}.

In this work, we employ Shannon wavelets \cite{cat08}, since they are regular and orthogonal functions with compact support in the Fourier domain. The regularity, as opposed to the Haar family used in \cite{Ortiz-Gracia2013}, gives us much better accuracy. The rapid decay of the scaling function in the Fourier domain replicates the behaviour of the ChF of the heavy-tail processes that we encounter in finance. 
A set of Shannon scaling functions $\phi_{m,k}$ in the subspace $\mathcal{V}_m$ is defined as,
\begin{equation}\label{eq:theta_sub}
				\phi_{m,k}(x) = 2^{m/2} \frac{\sin(\pi(2^m x-k))}{\pi(2^mx-k)} = 2^{m/2}\phi(2^mx-k), \quad k \in \mathbb{Z},
\end{equation}
			where,
\begin{equation*}
				\phi(z) = \sinc(z) = \left\lbrace
	            \begin{aligned}
                    & \frac{\sin(\pi z)}{\pi z}, & \text{if} \; z \neq 0,\\
                    & 1, & \text{if} \; z = 0,
	            \end{aligned}
        			\right.
\end{equation*}
is the scaling function, also called cardinal sine function.

The following lemma about the bound on the error of the orthogonal projection $\epsilon_m(x) := f(x) - P_m f(x)$ is provided in \cite{mar17}.

\begin{lem}[Lemma 3 of \cite{mar17}] \label{lem:projection_error} Let $\epsilon_m(x) := f(x) - P_m f(x)$ be the point-wise approximation error due to the projection of f into $V_m$. Then, 
\begin{equation}
|\epsilon_m(x)| \leq H(2^m \pi),
\end{equation}
where,
\begin{equation*}
H(\xi) := \dfrac{1}{2 \pi} \int_{|u| > \xi}\left|\hat{f}(u)\right| du,
\end{equation*}
is the normailzed mass of the two-side tails of $\hat{f}$.
\end{lem}

\subsection{SWIFT method}\label{sec:swiftmethod}
		
The SWIFT method belongs to the class of Fourier inversion methods for pricing European options within the discounted expected payoff formula (\ref{eq:integral_option_valuation}). The density function $f$ of (\ref{eq:integral_option_valuation}) is replaced by a finite combination of Shannon wavelets, and the coefficients of the approximation are computed from its ChF. The overall process provides an efficient way to obtain the value of an option, and it can be summarized, as in \cite{Ortiz-Gracia2016, mar17}, by a set of consecutive approximation steps, which are described below.

\begin{itemize} \label{swift_steps}

\item \textbf{Wavelet projection:} the function $f$ is replaced by its Shannon wavelet projection at scale $m \in \Z$,
\begin{equation} \label{step:wavelet_projection}
f(y|x) \approx f_1(y|x) := P_mf(y|x) = \sum_{k \in \Z} D_{m,k}(x) \phi_{m,k}(y),
\end{equation}
with $D_{m,k}(x):=\langle f(\cdot | x), \phi_{m,k} \rangle$.
\item \textbf{Series truncation:} the set of values of $k$ involved in the sum of expression (\ref{step:wavelet_projection}) is reduced to a finite interval $[k_1, k_2]$,
\begin{equation}
f_1(y|x) \approx f_2(y|x) = \sum_{k = k_1}^{k_2} D_{m,k}(x)\phi_{m,k}(y).
\end{equation}
It is important to notice that the first approximation lets us express,
\begin{equation}
f(2^{-m}k|x) \approx f_1(2^{-m}k|x) = 2^{m/2} D_{m,k}(x),
\end{equation}
which quickly justifies that, for any given $x$, the density coefficients vanish as $|k|$ increases, because the mass at the tails of $f$ tends to $0$. It is also worth noting that increasing $m$ will result in this mapping being less favorable. That is, for each $k$, $D_{m,k}$ will be bounded by a point closer to the center of the density function, potentially requiring to increase the range of values for $k$ in interval $[k_1, k_2]$.
\begin{remark}
From this point onward a symmetric interval $[1 - \eta, \eta]$ will be considered both for convenience and for consistency with the code implementation.
\end{remark}
\item \textbf{Density coefficients approximation:} the integral required to compute $D_{m,k}$ is replaced by an approximation $D_{m,k}^{*}$ as it will be shown in Section \ref{sec:sinc_integral},
\begin{equation}
f_2 (y|x) \approx f_3(y | x) = \sum_{k = 1 - \eta}^{\eta} D_{m,k}^{*}(x)\phi_{m,k}(y),
\end{equation}
We can then define $V_{m,k} := \int_{\R} \phi_{m,k}(y) v(y, T)dy$, and substitute $f_3(y|x)$ into expression (\ref{eq:integral_option_valuation}), obtaining,
\begin{equation*}
v(x,t) \approx v_3(x,t) := e^{-r \tau} \sum_{k = 1 - \eta}^{\eta} D_{m,k}^{*}(x)  V_{m,k}dy,
\end{equation*}

For European options, one can instead express it in terms of the strike-free payoff by defining $U_{m,k} := \dfrac{V_{m,k}}{K} = \int_{\R} \phi_{m,k}(y) g(y)dy$ obtaining,
\begin{equation}
v_3(x,t) := K e^{-r \tau} \sum_{k = 1 - \eta}^{\eta} D_{m,k}^{*}(x)  U_{m,k}dy.
\end{equation}

\item \textbf{Payoff coefficients approximation:} the integral required to compute $U_{m,k}$ is approximated in an analogous way as the integral to compute the density coefficients, and $U_{m,k}$ is replaced by an approximation $U_{m,k}^{*}$ as it will be shown in Section \ref{sec:sinc_integral},
\begin{equation}\label{eq:v_4}
v_3(x,t) \approx v_4(x,t) = e^{-r \tau} \sum_{k = 1 - \eta}^{\eta} D_{m,k}^{*}(x) U_{m,k}^{*}.
\end{equation}
These coefficients can be precomputed when initializing the SWIFT procedure and shared across different strikes and maturities, saving computation time.
\end{itemize}

\subsubsection{Density and payoff coefficients approximation}\label{sec:sinc_integral}
Density and payoff coefficients calculation rely on the approximation of $\sinc(x)$ by a finite combination of cosines (all the details can be found in \cite{Ortiz-Gracia2016} and \cite{mar17} and the references therein). As a result of this approximation, a new parameter called $J$ appears. This parameter will be labeled $J_d$ and $J_p$ to denote density and payoff coefficients, respectively.
Following the notation of \cite{mar17}, the density coefficients are given by,
\begin{equation}\label{eq:density_coeff}
D_{m,k}^{*} =  \dfrac{2^{m / 2}}{J_d} \sum_{j=1}^{J_d} \Re \left( \hat{f} \left(u_j^{d} 2^{m};x \right) e^{ik u_j^{d}} \right),
\end{equation}
where $\Re$ denotes de real part, $u_j^{d} = \dfrac{\pi}{2J_d}(2j - 1)$, and the payoff coefficients are given by,

$$U_{m,k} \approx U_{m,k}^{*}(-c, c) := \dfrac{2^{m/2}}{J_p}\sum_{j=1}^{J_p} \Re \left(e^{i k u_j^{p}} I_{j}(c)\right), $$
where,
\begin{equation} \label{eq:strike_free_int}
I_{j}(c) := \int_{|y| \leq c} g(y) e^{-i u_j^{p} y} dy = H(c) - H(-c), 
\end{equation}
and,
\begin{equation}
H(y) := -i e^{i u_j^{p} y} \left(\dfrac{1}{u_j^p} - \dfrac{e^y}{-i + u_j^p}\right), \quad u_j^{p} = \dfrac{\pi}{2J_p}(2j - 1).
\end{equation}

We can see that the value of $I_j(c)$ is periodic on $c$. In general, all the approximations to $\sinc(x)$ used in the SWIFT method are periodic, which can give rise to boundary issues and undervaluation of option prices when the option strikes approach to the boundary of $(-c, c)$. This issue also appears in the COS option pricing method \cite{Fang2008}, another Fourier-transform-based option pricing method closely rleated to the SWIFT method, and is discussed in \cite{mar17}. In that work, the authors use the independence between the parameter $c$ regulating the payoff integral domain and the parameter  $\eta$ regulating the wavelet series truncation, to carefully choose a value for $c$ to avoid this problem.
%
%
An iterative method to choose appropriate values for $m$, $\eta$, $J_d$, and $J_p$ is provided in \cite{Ortiz-Gracia2016,mar17}.

As most operators used in the SWIFT method are linear, one can easily obtain an expression for the option price gradient that will be used for calibration. In particular, the only dependence the European option price formula has on the model parameters vector $\boldsymbol{\theta}$ appears in the term $D_{m,k}^{*}$, so we can simply define,
\begin{equation}\label{eq:cmk_greeks}
D_{m,k}^{*(n)}(x; \boldsymbol{\theta}_i):= \dfrac{\partial^n D^*_{m,k}(x; \boldsymbol{\theta}_i)}{\partial\boldsymbol{\theta}_i^n} = \dfrac{2^{m / 2}}{J_d} \sum_{j=1}^{J_d} \Re\left(\frac{\partial^n \chf\left(u_j^d 2^{m}; x;\boldsymbol{\theta}_i\right)}{\partial\varsigma^n}e^{ik u_j^d}\right),
\end{equation}
and replace it into expression (\ref{eq:v_4}) to obtain the expression of the gradient.

\subsection{Speeding up the SWIFT method}
We elaborate on several enhancements of the SWIFT method in terms of efficiency, either on the pricing or during the calibration process. In Section \ref{subsec:fft}, it is explained how the density and payoff coefficients can be computed by means of the Fast Fourier Transform\footnote{In all the numerical examples in this article the C library FFTW will be used to compute the FFT \cite{fftw3}.} (FFT). Section \ref{subsec:multiple_strikes} is devoted to the adaptation of the SWIFT pricing machinery for multiple strikes valuation. Those two transformations where already pointed out in \cite{Ortiz-Gracia2016}. Moreover, when the vector of strikes meets a certain property, then the calibration can be carried out in a two stage procedure detailed in Section \ref{sec:fixedstrikes}. Finally, Section \ref{subsec:gradient} describes another improvement in regard to the option price gradient calculation.

\subsubsection{Fast computation of the density and payoff coefficients} \label{subsec:fft}
We start with a general expression of the summation term that appears in both the density and payoff coefficients approximation,
\begin{equation}
f_k = \sum_{j=1}^J g_j e^{i \dfrac{2j - 1}{2J}\pi k}
\end{equation}
We can extend it by defining $g_j = 0$ for $j = 0$ and $J < j < 2J$ and take the $j$-independent terms outside of the summation, obtaining,
\begin{equation}
f_k = e^{\dfrac{-i \pi k}{2J}}\sum_{j=0}^{2J - 1} g_j e^{i \dfrac{2j\pi k}{2J}}
\end{equation}
This last summation expression coincides with the Discrete Fourier Transform (DFT) of length $2J$ of $\{g_j\}$, and the computation of all the values $f_k$ can then be speeded up by the FFT implementation. 

\begin{remark}
Note that computing the density or payoff coefficients imposes a restriction on the wavelet series truncation parameter $2 \eta < J$.
\end{remark}

\subsubsection{Valuation with multiple strikes}\label{subsec:multiple_strikes}
A key property of the SWIFT method is that, given a scale of approximation $m$, the payoff and density coefficient associated to each wavelet $\phi_{m,k}$ can be computed through two FFTs (one for all the density coefficients, and one for all payoff coefficients). Without this property, the SWIFT computation speed would not be competitive with other numerical option pricing methods \cite{flo20}.

In the option calibration problem, one usually needs to consider the option prices of several options at different strikes. In this specific case, if one were to compute the option prike of $M$ options at strikes $\boldsymbol{K} := (K_1, \ldots, K_M)^{T}$, then the formulation proposed in expression (\ref{eq:v_4}) would need to recompute the density and payoffs coefficients for every strike $K_i$. This involves evaluating the ChF $\eta \cdot J_p \cdot M$ times, an operation which, for the Heston model is more costly than evaluating the strike-free payoff function, or its integral. As stated in \cite{Ortiz-Gracia2016} one can improve the computation time of option pricing for multiple strikes when $\hat{f}(u; x) = \hat{f}(u) e^{-iux}$, a property present in both L\'evy and Heston models.

As stated in \cite{mar17}, let us start from expression (\ref{eq:v_4}), and considering the previously mentioned vector of strikes $\boldsymbol{K}$, with its associate vector of initial states $\boldsymbol{x} := (\log(S_0/K_1), \ldots, \log(S_0/K_M))^{T}$. We can then substitute the density coefficients approximation (\ref{eq:density_coeff}) into the option price expression (\ref{eq:v_4}) and interchange the two resulting summations, obtaining,

\begin{align}
v_4(\boldsymbol{x}, t) :&=  e^{-r\tau} \boldsymbol{K} \sum_{k = 1 - \eta}^{\eta} \Re \left\{ \sum_{j=1}^{J_d} \hat{f}(u_j^d 2^m; \boldsymbol{x}) e^{i u_j^{d} k} U_{m,k}^{*}(-c,c) \right\} \\
& =e^{-r\tau}\boldsymbol{K} \sum_{j=1}^{J_d} \Re \left\{ \hat{f}(u_j^{d} 2^m;\boldsymbol{x}) \left[ \sum_{k = 1-\eta}^{\eta} U_{m,k}^{*}(-c,c) e^{i u_j^{d} k} \right] \right\} \\
&= e^{-r\tau}\boldsymbol{K} \sum_{j=1}^{J_d} \Re \left\{ \hat{f}(u_j^{d} 2^m) e^{-i u_j^{d}2^m \boldsymbol{x}} \left[ \sum_{k = 1-\eta}^{\eta} U_{m,k}^{*}(-c,c) e^{i u_j^{d} k} \right] \right\} \\
\label{step:alternative_final}
&= e^{-r\tau}\boldsymbol{K} \sum_{j=1}^{J_d} \Re \left\{ \hat{f}(u_j^{d} 2^m) e^{-i u_j^{d} 2^m \boldsymbol{x}} \tilde{U}_j(-c,c) \right\}.
\end{align}

where, $$ \tilde{U}_j(-c,c) := \sum_{k = 1-\eta}^{\eta} U_{m,k}^{*}(-c,c) e^{i u_j^{d} k}.$$

The original formulation from expression (\ref{eq:v_4}) requires the following computations,
\begin{itemize}
\item For each of the M strikes:
\begin{itemize}
\item 1 FFT of length $2J_d$ to compute $2\eta$ density coefficients.
\item $J_d$ evaluations of the ChF $\hat{f}(u_j^d 2^m;x)$.
\end{itemize}
\item 1 FFT of length $2J_p$ to compute $2\eta$ payoff coefficients.
\item $J_p$ evaluations of the strike-free payoff integral $I_j(c)$ defined in expression (\ref{eq:strike_free_int}).
\end{itemize}
The payoff computations are independent of the strike price and can be computed only once, and be reused for all strikes.

On the other hand, the alternative formulation provided in expression (\ref{step:alternative_final}) requires,
\begin{itemize}
\item For each of the M strikes:
\begin{itemize}
\item $J_d$ evaluations of the $x$-dependent term of the ChF $e^{-iu_j^d \boldsymbol{x}}$.
\end{itemize}
\item $J_d$ evaluations of the $x$-independent ChF $\hat{f}(u_j^d 2^m)$.
\item 2 FFT of lengths $2 \eta$ and $2J_p$ to compute the $J_d$ values of $\tilde{U}_j(-c,c)$.
\item $J_p$ evaluations of the strike-free payoff integral $I_j(c)$ defined in expression (\ref{eq:strike_free_int}).
\end{itemize}
The required values of $\hat{f}(u_j^d 2^m)$, $\tilde{U}_j(-c,c)$, and the values of $I_j(c)$ required to compute the latter, can be computed only once and be reused for all strikes.
\begin{remark}
In general, whenever the dependency on $x$ in $\hat{f}(u; x)$ can be easily isolated and is cheap to compute, one can benefit from the alternative formulation proposed in this section.
\end{remark}

The computation of the ChF tends to be more expensive than the computation of the payoff integral, so a SWIFT implementation through expression (\ref{step:alternative_final}) tends to outperform the one through expression (\ref{eq:v_4}) when several strike prices are involved. A discussion on the benefits of using a formulation equivalent to the one provided in (\ref{step:alternative_final}) for multiple strikes appears in \cite{Ortiz-Gracia2016} where it is shown that is possible to define $F_j := \hat{f}(u_j^d 2^m)$ and compute in advance its $J_d$ required values once and reuse them for all strikes.

\subsubsection{Fixed set of strikes}\label{sec:fixedstrikes}
Let us consider,
\begin{equation} 
G_j := \left\{ \begin{array}{rcl} F_j, \tilde{U}(-c, c) & \mbox{for} & j \leq J_d, \\
 0, & \mbox{for} & J_d < j \leq 2 J_d. \end{array}\right.
\end{equation}
Then expression (\ref{step:alternative_final}) can be rearranged as,
\begin{equation}\label{step:multiple_strikes1}
v_4(x, t) = e^{-r\tau}\boldsymbol{K} \Re \left\{e^{\dfrac{\pi i 2^m \boldsymbol{x}}{J_d}}. \sum_{j=1}^{2J_d} G_j e^{- \dfrac{2\pi i j 2^m \boldsymbol{x}}{2J_d}} \right\}
\end{equation}
If one chooses selectively the values of the strikes $K_l$, so that $2^m x_l$ is an integer number, this computation can be speeded up by the use of an FFT algorithm. If one chooses $x_k := \dfrac{2k - J_d}{2^{m+1}}$, then expression (\ref{step:multiple_strikes1}) becomes,
\begin{equation}\label{step:multiple_strikes2}
v_4(x, t)  = e^{-r\tau}\boldsymbol{K} Re \left\{e^{\dfrac{\pi i 2^m \boldsymbol{x}}{J_d}} \sum_{j=1}^{2J_d} G_j e^{- \dfrac{2\pi i j k}{2J_d}} e^{\pi i j}  \right\} 
 = e^{-r\tau}\boldsymbol{K} Re \left\{e^{\dfrac{\pi i 2^m \boldsymbol{x}}{J_d}} \sum_{j=1}^{2J_d} \tilde{G_j} e^{- \dfrac{2\pi i j k}{2J_d}} \right\}, 
\end{equation}
where, 
\begin{equation}
\tilde{G}_j := G_j e^{\pi i j} = G_j (-1)^j.
\end{equation}
\begin{remark}
Note that, as with other FFT-based computations presented in this work, this approach imposes a bound $M \leq J_d$ on the number of different strikes that can be computed with the FFT.
\end{remark}
If one considers the domain $\mathcal{D}$ of $x = \log(S_t/K)$, this approach allows pricing options in a symmetrical boundary $(-\dfrac{J_d}{2^{m+1}},\dfrac{J_d}{2^{m+1}}) \in \mathcal{D}$ at $J_d$ uniformly distributed points at distance $2^m$.
We can not usually choose the strike prices at which to price the options, particularly not when calibrating a model with real market data, as only a limited set of strike values are listed on any exchange market, but this method could be used to quickly compute the option prices of an already calibrated model at a grid of points that could be tuned by the choice of $m$ and $J_d$. Then the option prices at any intermediate strike could be interpolated with the help of a derivative-free spline (or, if the derivative with respect to K in expression (\ref{step:multiple_strikes2}) preserves the same speed properties, with the help of any spline method that uses derivatives).

\subsubsection{Option price gradient}\label{subsec:gradient}
The option price gradient must be computed during the calibration process that will be presented in Section \ref{chap:optimization_problem}. All the aforementioned techniques can be applied to the option price gradient. We can also enumerate three more speed up properties,

\begin{itemize}
\item The value of $e^{-i u_j^d 2^m x_l}$ can be reused for the price as well as for the gradient computations.
\item If the parameters of the SWIFT method, are not changed during the gradient descent used in the calibration problem, then the values of both $\tilde{U}_j(-c,c)$ and $e^{-i u_j^d 2^m x_l}$ can be reused throughout all the calibration steps.
\item We can reuse the values of $\hat{f}(u_j^d 2^m)$ from the price computation to compute the gradient. 
\end{itemize}
So combining all the speed properties above, when solving a gradient-based calibration problem, we only need to first compute $\tilde{U}_j(-c,c)$ and $e^{-i u_j^d 2^m x_l}$, and then in each gradient-descent step, one can simultaneously calculate both the price and the gradient of all strikes by computing once for each $j \in [1, J_d]$ the values of $\hat{f}(u_j^d 2^m)$ and $\boldsymbol{h}(u_j^d 2^m)$.

\section{Calibration} \label{chap:optimization_problem}
Calibrating a model for the asset underlying the option, is a sophisticated procedure that requires highly efficient numerical methods. In particular, the pricing of the options used for calibration should be carried out by means of an accurate, fast and robust valuation method. In this work, we calibrate the Heston model by means of the SWIFT method, and compare it with the Heston's pricing method of \cite{cui17} that we have written in a more compact form in Lemma \ref{lem:Hestonpricing}. 
It is worth noting that the choice of a specific objective function will have an impact on how accurately the model of the underlying asset that we will obtain through callibration will describe different real market scenarios \cite{chr02}. For comparison sake, the same one as in \cite{cui17} will be used.

Let $V^*(x_i, \tau_i)$ be the market price of a European call option and
 $V(\boldsymbol{\theta}; x_i, \tau_i)$ be the price at the same strikes and maturities obtained by using either the SWIFT method or the Heston pricing formula in expression (\ref{eq:heston_analytic}) of Lemma \ref{lem:Hestonpricing}. Let us also assume that we use a set of $n$ different options to calibrate the model, so that $i \in [1,n] \subset \Z$. Then,
the calibration of the model is defined as the minimization problem, 
\begin{align}
\label{eq:opt_problem}
& min_{\boldsymbol{\theta} \in \R^5} f(\boldsymbol{\theta}), & f(\boldsymbol{\theta}) := \dfrac{1}{2}||\boldsymbol{r}(\theta)||^2 = \dfrac{1}{2} \boldsymbol{r}^T(\boldsymbol{\theta})\boldsymbol{r}(\boldsymbol{\theta}), &
\end{align}
where $\boldsymbol{r}(\boldsymbol{\theta})$ is the $n$-dimensional vector of the residuals obtained when pricing the options considered for calibration using the model parameters. That is,
\begin{align} \label{eq:opt_problem2}
\boldsymbol{r}(\boldsymbol{\theta}) := \left[r_1(\boldsymbol{\theta}), \ldots, r_n(\boldsymbol{\theta}) \right]^T, 
&& r_i(\boldsymbol{\theta}) :=  V(\boldsymbol{\theta}; x_i, \tau_i) - V^*(x_i, \tau_i), && i = 1, \ldots, n.
\end{align}
If we calculate the Jacobian of $\boldsymbol{r}$, gives us,
\begin{equation}
\boldsymbol{J} := \nabla_{\boldsymbol{\theta}} \boldsymbol{r}^T = \nabla_{\boldsymbol{\theta}} V(\boldsymbol{\theta}; \boldsymbol{x}, \tau),
\end{equation}
where,
\begin{equation}
J_{ji} = \left(\dfrac{\partial r_i}{\partial \theta_j} \right) = \left(\dfrac{\partial V(\boldsymbol{\theta}; x_i, \tau_i)}{\partial \theta_j} \right).
\end{equation}
The Hessian matrix of the residual element $r_i$ reads,
\begin{equation}
\boldsymbol{H}(r_i) := \nabla_{\boldsymbol{\theta}} \nabla_{\boldsymbol{\theta}}^{T} r_i = \nabla_{\boldsymbol{\theta}} \nabla_{\boldsymbol{\theta}}^{T} V(\boldsymbol{\theta}; x_i, \tau_i),
\end{equation}
where,
\begin{equation}
H_{jk}(r_i) = \left(\dfrac{\partial^2 r_i}{\partial \theta_j \partial \theta_k} \right) = \left(\dfrac{\partial^2 V(\boldsymbol{\theta}; x_i, \tau_i)}{\partial \theta_j\partial \theta_k} \right).
\end{equation}
Then, the gradient and Hessian matrix of the objective function defined by expressions (\ref{eq:opt_problem}) and (\ref{eq:opt_problem2}) are,
\begin{align}
\nabla_{\boldsymbol{\theta}}f(\boldsymbol{\theta}) &= \boldsymbol{J}\boldsymbol{r}, \\
\nabla_{\boldsymbol{\theta}}\nabla_{\boldsymbol{\theta}}^{T} f(\boldsymbol{\theta}) &= \boldsymbol{J}\boldsymbol{J}^{T} + \sum_{i = 1}^{M} r_i \boldsymbol{H}(r_i).
\end{align}
We solve the optimization problem (\ref{eq:opt_problem}) by means of the LM\footnote{As for the implementation, we use the LEVMAR C package \cite{levmar} as well as the LAPACK linear algebra package \cite{lapack}.} method. This method is as a blend of gradient descent (GD) and Gauss-Newton (GN) iteration, depending on whether the current guess is close or far from the optimum. The exact expression of the step $\Delta \boldsymbol{\theta}$ is,
\begin{equation}
\Delta \boldsymbol{\theta} = (\boldsymbol{J}\boldsymbol{J}^{T} + \mu \boldsymbol{I})^{-1} \nabla_{\boldsymbol{\theta}} f(\boldsymbol{\theta}),
\end{equation}
where $\boldsymbol{I}$ is the identity matrix and $\boldsymbol{J}\boldsymbol{J}^{T} + \mu \boldsymbol{I}$ substitutes the Hessian matrix used in the Newton method.
When the current guess is far from the optimum, a large value is given to $\mu$ so that,
\begin{equation}
\Delta \boldsymbol{\theta} \approx \Delta \boldsymbol{\theta}^{(\text{SD})} := (\mu \boldsymbol{I})^{-1} \nabla_{\boldsymbol{\theta}} f(\boldsymbol{\theta}),
\end{equation}
and a small step of a steepest-descent method is taken.
When the current guess is close to the optimum, a small value is given to $\mu$ so that,
\begin{equation}
\Delta \boldsymbol{\theta} \approx \Delta \boldsymbol{\theta}^{(\text{GN})} := (\boldsymbol{J}\boldsymbol{J}^{T})^{-1} \nabla_{\boldsymbol{\theta}} f(\boldsymbol{\theta}),
\end{equation}
and the Hessian usually used in the Newton method is replaced by its GN approximation. This approximation is reliable if either $r_i$ or $\boldsymbol{H}(r_i)$ are small, and \cite{cui17} justifies its usage by conjecturing that $f$ is nearly linear, a condition that guarantees the latter. We should note that even if $f$ were not linear, then LM should only use small values of $\mu$ when $|\boldsymbol{r}|$ is small at the current step of the optimization problem.

The iterative algorithm implemented for the LM method stops when, at a certain iteration $n$, any of the following criteria is fulfilled,
\begin{align}
&|\boldsymbol{r}_n| \leq \epsilon_1, \label{eq:firstcriteria}\\
&|\boldsymbol{J}_n|_{\infty} \leq \epsilon_2, \label{eq:secondcriteria} \\
&\dfrac{|\nabla \boldsymbol{\theta}_n|}{|\boldsymbol{\theta}_n|} \leq \epsilon_3. \label{eq:thirdcriteria}
\end{align}

The first stopping criteria (\ref{eq:firstcriteria}) is fulfilled when the objective function defined by expressions (\ref{eq:opt_problem}) and (\ref{eq:opt_problem2}) has reached a value closer to zero than the prescribed tolerance. It is only when the method stops due to this criteria that we will consider that the model has been properly calibrated. The second criteria (\ref{eq:secondcriteria}) corresponds to a flat gradient, and the third (\ref{eq:thirdcriteria}) corresponds to a stagnating update (this last one has never happened while testing the convergence during the Heston model calibration).

\section{Numerical results} \label{chapter:study}
In this word, the SWIFT\footnote{The code implemented for this work can be consulted in the following Github public repository:
https://https://github.com/eudaldrg/SWIFTOptionCalibration.} method is used to calibrate a Heston model with European call options price data at different strikes and maturities, and it will be compared to the pricing and calibration method based on expression (\ref{eq:heston_analytic}) proposed by \cite{cui17}, which for the sake of readability will be called Cui pricer (CP). CP will be implemented using a Gauss-Legendre quadrature with 64 nodes for its numerical integration step. The upper limit of the integral will be truncated, whenever possible, at $\overline{u} = 200$, but will be adjusted if necessary. The calibration process will consist of applying an LM method to the objective function defined in expression (\ref{eq:opt_problem}). The SWIFT method will be implemented using the ChF expression and its derivatives provided also in \cite{cui17}. We perform a wide variety of tests that can be summarized as follows,
\begin{itemize}
\item \textbf{Stress tests:}
the CP and SWIFT methods will be tested with several combinations of extreme strikes (ATM and deep ITM and OTM) as well as with long-term and short-term maturities to detect any possible limitation or numerical issue in a wide usage range.
\item \textbf{Speed\footnote{The computations were performed on a 64-bit Ubuntu 18.04.4 LTS with a 3.70GHz Intel Core i7-8700K processor and 62.8 gigabytes of ram.} tests:}
the option calibration speeds for the regular SWIFT method (defined by expression (\ref{eq:v_4})) and the one devised to quickly compute several option prices with different strike and same maturity (defined by expression (\ref{step:alternative_final}), which will be denoted KSWIFT), will be compared against CP for three different strike and expiry sets to check whether the multiple-strike alternative formulation is necessary to obtain a competitive option calibration method. These scenarios will represent,
\begin{itemize}
\item A single expiry and multiple strikes.
\item A fixed number of maturities and a fixed number of strikes per maturity.
\item Different expiries for each strike.
\end{itemize}
When computing options with more than one different strike, a combination of OTM and ITM options will be used to provide an heterogeneous sample of contracts. Similarly, when more than one maturity is considered, a sample of long- and short-term expiries will be used.




\item \textbf{Realistic convergence tests:}
as in \cite{cui17}, convergence of the method will be tested for realistic model parameters representative of long-dated Foreign Exchange (FX), interest rate, and equity options, as they are relevant and, according to \cite{gla11},challenging for simulations of the Heson model.






\end{itemize}
Several sets of Heston parameters will be used for the different numerical tests and are presented in Table \ref{table:heston_params}. The last three sets of parameters are representative of long-term FX, interest rate, and equity options respectively \cite{and08}.

\begin{table}[h]
\begin{center}
 \begin{tabular}{|c | c | c | c | c | c |} 
 \hline
 Name & $\kappa$ & $\overline{\nu}$ & $\sigma$ & $\rho$ & $\nu_0$ \\ [0.5ex] 
 \hline
 $\boldsymbol{\theta}^{(1)}$ & 3 & 0.1 & 0.25 & -0.8 & 0.08 \\ 
 \hline
 $\boldsymbol{\theta}^{(2)}$ & 1.5768 & 0.0398 & 0.0175 & -0.5711 & 0.0175 \\ 
 \hline
 $\boldsymbol{\theta}^{(\text{FX})}$ & 0.5 & 0.04 & 1 & -0.9 & 0.04 \\ 
 \hline
 $\boldsymbol{\theta}^{(\text{IR})}$ & 0.3 & 0.04 & 0.9 & -0.5 & 0.04 \\ 
 \hline
 $\boldsymbol{\theta}^{(\text{EQ})}$ & 1 & 0.09 & 1 & 0.04 & 0.09 \\ 
 \hline
\end{tabular}
\end{center}
\caption{Set of Heston parameters used in the numerical tests.}\label{table:heston_params}
\end{table}
\begin{remark}
$\boldsymbol{\theta}^{(1)}$ is obtained from \cite{cui17} and $\boldsymbol{\theta}^{(2)}$ is a plausible set of parameters proposed by us. It may not be representative of any real world market, but it is only used as our objective value in our speed tests. In Section \ref{subsec:speed}, we will use an initial guess of $\boldsymbol{\theta}^{(2)}_{0} = (1.5768, 0.0398, 0.5751, -0.5711, 0.0175)^{T}$.
\end{remark}

\subsection{Stress tests}\label{subsec:stress}

Deep ITM and OTM call options are priced together with ATM call options for long- and short-term maturities using the SWIFT method at two different scales of approximation ($m = 3$ and $m = 7$) and the CP method. The time until maturity $\tau$ is given in years. Thus, the expiries of 0.04 attempt to simulate a situation of around two weeks until expiration of the option contract.

\begin{table}[!h]
\begin{center}
 \begin{tabular}{|c | c | c | c | c | c | c |} 
 \hline
 Parameters & S & K & $\tau$ & $V_{\text{SW}}^{3}$ & $V_{\text{SW}}^{7}$ & 
  $V_{\text{CP}}$
 \\ [0.5ex] 
 \hline
 $\boldsymbol{\theta}^{(1)}$ & 100 & 50  & 45   & 65.565 & nan & nan \\ 
 \hline
 $\boldsymbol{\theta}^{(1)}$ & 100 & 100 & 45   & 46.911 & nan & nan \\ 
 \hline
 $\boldsymbol{\theta}^{(1)}$ & 100 & 200 & 45   & 27.198 & nan & nan \\ 
 \hline
 $\boldsymbol{\theta}^{(1)}$ & 100 & 50  & 0.04 & 44.221 & 50.000 & 50.000 \\ 
 \hline
 $\boldsymbol{\theta}^{(1)}$ & 100 & 100 & 0.04 & 0.380  & 1.045 & 1.046 \\ 
 \hline
 $\boldsymbol{\theta}^{(1)}$ & 100 & 200 & 0.04 & 0      & 0 & 1.079e-3 \\
 \hline
\end{tabular}
\end{center}
\caption{Set of Heston parameters used in the numerical tests. }\label{table:stress}
\end{table}
Table \ref{table:stress} presents the pricing results. We call $V_{\text{SW}}^{3}$ and $V_{\text{SW}}^{7}$ to the prices given by the SWIFT method at scales of approximation $m=3$ and $m=5$, respectively, while $V_{\text{CP}}$ refers to the price obtained with the CP method. Both methods run into numerical issues with either extremely large or extremely small expiries provided no other changes are performed in the methods.

\begin{itemize}
\item CP and $V_{\text{SW}}^{7}$ produced not a number (nan) results when evaluating very long expiries. Looking at the option price execution with the integrated debugger of the GDB compiler \cite{wil16} showed that expression (\ref{eq:char_cui}) runs into numerical overflow when the exponent $\frac{d \tau}{2}$ of its hyperbolic functions is big enough (the same error happened when using the original expression provided in \cite{rol10}). In most of the tests above, the overflow could be avoided when carefully setting an appropriate value for the upper bound $\overline{u}$ of integral in expression (\ref{eq:heston_analytic}), and by using a smaller value of the scale $m$. The error can also be avoided by selecting the ChF expression provided in \cite{sch04} (we use it later on, and we denote the obtained prices by $V_{\text{SH}}$ and present the results in Table \ref{table:second_attempt}). 



\item The SWIFT method at scale $m = 3$ tends to underprice short expiry options. After checking the SWIFT parameters obtained through the parameter choice method defined in Section \ref{sec:swift}, it was observed that the initial value for $\eta$, obtained by simply using the cumulant expression proposed in \cite{Ortiz-Gracia2016}, resulted in a truncated Shannon wavelet expansion that did not cover a sufficient domain of the density function $f(y | x)$. A dynamical choice of the parameter $\eta$ based on the calculation of the area underneath the curve of the density function, as described in \cite{Ortiz-Gracia2016}, can avoid this issue.
Increasing the value of $m$ also fixes the problem.

\item None of the methods can handle the deep OTM option with a short expiry. The expected value should be close to but bigger than 0, as there are only 10 trading days to expiry and the price of the underlying should increase $50\%$ so that the option contract would not expire worthless. CP value seems too high and, in fact, when moving the value of $\overline{u}$ in the interval $[100, 400]$ the price never clearly converges to a certain value, and it can give higher estimates for $\overline{u} > 200$ than $1.079e-3$, or even negative values. Changing the ChF expression does not fix this issue. SWIFT consistently gives it a price of 0. The contribution that makes the price different than zero probably lies on the tails of the distribution function, and one would require a really big value of $c$ so that a point with a positive payoff is even considered in expression (\ref{step:alternative_final}).
\end{itemize}
Table \ref{table:second_attempt} shows some results either selecting an appropriate value $\overline{u}$ or keeping $\overline{u}$ equal to 200 and using the ChF from \cite{sch04}. The column $\overline{u}$ indicates the value at which CP integral is truncated, and the option price obtained is shown in column $V_{\text{CP}}$. Column $V_{\text{SH}}$ shows the price obtained when keeping $\overline{u} = 200$ but implementing CP using the ChF provided in \cite{sch04}. The last row is an example of the negative values obtained in the deep OTM short-term call.

\begin{table}[!h]
\begin{center}
 \begin{tabular}{|c | c | c | c | c | c | c |} 
 \hline
 Parameters & S & K & $\tau$ & $\overline{u}$ & $V_{\text{CP}}$ & $V_{\text{SH}}$ \\ [0.5ex] 
 \hline
 $\boldsymbol{\theta}^{(1)}$ & 100 & 50  & 45  & 6 & 65.565 & 65.565 \\ 
 \hline
 $\boldsymbol{\theta}^{(1)}$ & 100 & 100 & 45   & 6 & 46.911 & 46.911 \\ 
 \hline
 $\boldsymbol{\theta}^{(1)}$ & 100 & 200 & 45   & 6 & 27.198 & 27.198 \\ 
 \hline
 $\boldsymbol{\theta}^{(1)}$ & 100 & 50  & 0.04 & 200 & 50.000 & 50.000 \\ 
 \hline
 $\boldsymbol{\theta}^{(1)}$ & 100 & 100 & 0.04 & 200 & 1.046 & 1.046 \\ 
 \hline
 $\boldsymbol{\theta}^{(1)}$ & 100 & 200 & 0.04 & 300 & -1.174e-5 & 1.079e-3 \\
 \hline
\end{tabular}
\end{center}
\caption{Results for different $\overline{u}$ and/or using the ChF from \cite{sch04}.}\label{table:second_attempt}
\end{table}
As it has been shown so far, a crucial step of the calibracion process is the selection of $\overline{u}$ for the CP method and $m$ for the SWIFT method. A method to set an optimal value $\overline{u}$ is not provided in \cite{cui17}, and it is therefore a matter of trial and error, since it must be manually determined when changing the time to expiry of the options one wants to price. As for the SWIFT method, we can use the iterative parameter choice provided in \cite{mar17}. In particular, a suitable scale of approximation $m$ can be selected by means of Lemma \ref{lem:projection_error}. Once the level of approximation $m$ is fixed, the parameter $\eta$ can be adaptively calculated in order to determine the wavelet series truncation accurately.

\subsection{Speed tests}\label{subsec:speed}
The calibration speed has been tested for three different sets of strike prices and maturities, which are available in Appendix \ref{app:strikes}. In order to be sure that the calibration problem was properly converging, we chose $\boldsymbol{\theta}^{(2)}$ as  the objective value for the Heston parameters. When testing each set we perform the following actions,

\begin{itemize}
\item Generate option price values for each strike-expiry pair using $\boldsymbol{\theta}^{(2)}$ as input. For this step, we used SWIFT method (we also generated them using CP method and checked that the difference between both results stayed under $10^{-7}$).
\item Chose an initial guess for the calibration problem $\boldsymbol{\theta}^{(2)}_0$.
\item Solve the calibration problem with the desired method using $\boldsymbol{\theta}^{(2)}_0$ as initial guess and use as inputs the strike-expiry pairs and the prices obtained in the first step.
\end{itemize}

Set 2 has been obtained from the code provided in \cite{cui17}, and represents a total of 40 options, distributed in 8 different maturities, each maturity with 5 different strikes. Set 1 and set 3 are extreme cases derived from set 2, in order to test the best and worst calibration points distribution for KSWIFT. Set 1 has the same 40 different strike prices than set 2, but only a single maturity. What we denote as set 3 is not really a different data set, since it consists of running the calibration problem with set 2 and preventing the KSWIFT algorithm from applying the speed up techniques discussed in Section \ref{subsec:multiple_strikes}. For this reason, in Table \ref{table:speed} set 3 contains values only for KSWIFT (as set 3 is equivalent to set 2 for the other two calibration methods).

\begin{table}[!h]
\begin{center}
 \begin{tabular}{|c | c | c | c | c | c |} 
 \hline
 Strike and maturities set & Heston parameters & Method & Time (seconds) & I & $\epsilon_1$ \\ [0.5ex] 
 \hline
 Set 1 & $\boldsymbol{\theta}^{(2)}$ & SWIFT & 6.9 & 10 & 3.932e-11 \\ 
 \hline
 Set 1 & $\boldsymbol{\theta}^{(2)}$ & KSWIFT & 4.5e-3 & 10 & 3.932e-11 \\
 \hline
 Set 1 & $\boldsymbol{\theta}^{(2)}$ & CP & 4.6e-2 & 10 & 3.932e-11 \\
 \hline
 Set 2 & $\boldsymbol{\theta}^{(2)}$ & SWIFT & 35.9 & 13 & 1.002e-12 \\
 \hline
 Set 2 & $\boldsymbol{\theta}^{(2)}$ & KSWIFT & 5.0e-2 & 13 & 1.002e-12 \\
 \hline
 Set 2 & $\boldsymbol{\theta}^{(2)}$ & CP & 6.3e-2 & 13 & 1.002e-12 \\
 \hline
 Set 3 & $\boldsymbol{\theta}^{(2)}$ & KSWIFT  & 1.7e-1 & 13 & 1.002e-12\\
 \hline
\end{tabular}
\caption{Iterations, time needed to calibrate each speed scenario and objective function value reached. I refers to the number of iterations that LM requieres until it stops, and $\epsilon_1$ corresponds to LM first stopping criteria (see Section \ref{chap:optimization_problem}), which refers to the objective function final value.}\label{table:speed}
\end{center}
\end{table}

The values for KSWIFT and CP have been averaged over 100 executions of the calibration to provide a good estimate of the required calibration time. It can be seen that regular SWIFT is orders of magnitude slower without averaging the required time over several executions. Hence, the multiple-strike alternative formulation presented in Section \ref{subsec:multiple_strikes} and all the speed-up techniques discussed through Section \ref{sec:swift} are necessary to provide a competitive method that can be used for real-time model updating.

KSWIFT performance is comparable to CP for set 2, an order of magnitude faster for set 1, and an order of magnitude slower for set 3. It can be argued that both set 1 and set 3 are extreme cases that are not really relevant for real option trading situations. One would rarely use a single strike per expiry to calibrate an option pricing model, and using data from a single expiry only seems reasonable when trading a single option expiry (in this case, one could benefit from the speed properties of KSWIFT on scenarios like set 1). According to \cite{cui17}, a reasonable calibration scenario consists of using option prices from strikes at $0\%$, $\pm 25\%$, and $\pm 50\%$ BS delta (derivative of the option price with respect to the underlying price value. It has a closed analytical expression for European BS options).

The calibration time of KSWIFT and CP is about $0.05$ seconds, which seems sufficient for real-time model updating to provide market information to a human trader. In a more computationally demanding trading environments, like high-frequency trading neither KSWIFT nor CP would be competitive enough.

\begin{remark}
All the single expiry tests (the first three tests on Table \ref{table:speed}) converged to an approximated value different than $\boldsymbol{\theta}^{(2)}$ but approximated all the option prices properly. Using different initial guesses lead to different approximated values which minimized the objective function. It would be interesting to see whether this is a property of the Heston distribution (that is, it has at least a degree of freedom when defined from option prices in a single expiry) or it is due to the specific scenario being tested.
\end{remark}

\subsection{Realistic convergence tests}
We use the same procedure as in previous section to solve several different calibration problems. For each case, we use one of the proposed realistic parameter sets as objective value and generate option prices for each strike-expiry pair. Then, for each objective parameter set, 100 different initial guesses are generated. Each component of the initial parameters guess is drawn uniformly and randomly within $\pm 10\%$ distance of the optimal value.
According to \cite{cui17} this is representative of real option calibration as, usually, the initial guess used for a certain calibration problem is the last available parameter estimation. If the calibration is updated fast enough, it is expected that the initial guess will be this close to the optimum. The maturities used in \cite{cui17} are not available, so for these tests the strike-expiry set 2 will be used.
\begin{table}[!h]
\begin{center}
 \begin{tabular}{|c | c | c | c|} 
 \hline
  & $\boldsymbol{\theta}^{(\text{FX})}$ &$\boldsymbol{\theta}^{(\text{IR})}$ & $\boldsymbol{\theta}^{(\text{EQ})}$ \\ [0.5ex] 
 \hline
 $|\kappa^{a} - \kappa^{*}|$ & 6.640e-4 & 2.657e-4 & 1.160e-3 \\ 
 \hline
 $|\overline{\nu}^{a} - \overline{\nu}^{*}|$ & 1.547e-4 & 1.321e-5 & 1.746e-5\\
 \hline
 $|\sigma^{a} - \sigma^{*}|$ & 1.978e-3 & 2.248e-4 & 3.725e-4 \\
 \hline
 $|\rho^{a} - \rho^{*}|$ & 2.649e-4 & 1.365e-5 & 8.661e-6  \\
 \hline
 $|\nu_0^{a} - \nu_0^{*}|$ & 3.629e-5 & 4.790e-6 & 8.339e-6  \\
 \hline
 Iterations & 14 & 6 & 7 \\
 \hline
 Time (seconds) & 3.3e-1 & 1.9e-1 & 2.0e-1 \\
 \hline
 $\epsilon_1$ & 2.867e-11 & 2.030e-11 & 3.643e-11\\
 \hline
\end{tabular}
\caption{Convergence statistics averaged over 100 calibrations. $x^a$ refers to the calibration problem's estimation of variable $x^a$. For example $|\kappa^{a} - \kappa^{*}|$ refers to how close LM approximation of $\kappa$ was to the real value.}\label{table:realistic}
\end{center}
\end{table}
As can be seen in Table \ref{table:realistic}, even under challenging parameters setups representative of real option trading, KSWIFT is able to provide accurate estimations of the Heston parameters, taking on average a computation time of hundreds of milliseconds. These results, both in terms of speed as in terms of accuracy, are comparable to the tests in \cite{cui17}, so it is concluded that KSWIFT is as efficient as CP for real market scenarios. Further, if we take into account the robustness of KSWIFT in terms of the a priori knowledge of its parameters, as stated in Section \ref{sec:swiftmethod}, we conclude that KSWIFT is a very competitive method for calibration.

\section{Conclusions and future research} \label{chap:conclusions}
We have investigated the problem of calibrating the Heston model, which belongs to the class of stochastic volatility models. An extension of the SWIFT method has been provided in this work for European options calibration, along with novel speed-up techniques, which can radically improve the performance when several of the priced and calibrated options have the same time to maturity.

Some numerical issues arise with the ChF for very long-term expiries. Following the a priori knowledge of parameters selection for the SWIFT method seems to be enough to avoid these problems, while the parameters of CP need to be adjusted manually.
The proposed speed-up techniques are deemed necessary in order to make SWIFT a competitive calibration method, as it has been seen in the numerical speed tests. In particular, it has been shown that the only situation where the proposed calibration is significantly slower than CP is when one calibrates the model with many different maturities with no more than one or two strikes per maturity. As the number of strikes per expiry increases, the relative speed of the SWIFT method increases, and it is about ten times faster than CP when calibrating 40 options with a single maturity. Both extreme situations are not representative of real option trading needs, and for a reasonably real situation of 5 strikes per expiry, the SWIFT technique is slightly faster than CP. A SWIFT implementation without the previously discussed speed-up techniques has also been tested and deemed non-competitive, with calibration times that reached the dozens of seconds.
Further, the proposed calibration strategy passes the realistic calibration tests for challenging Heston model parameters setups presented.

In summary, the proposed SWIFT method is a robust and efficient machinery for real-time updating of option models used in human-supervised trading schemes. Neither SWIFT nor CP method are suitable for the most demanding algorithmic trading situations, like high-frequency trading. Several future work may encompass the following topics,
\begin{itemize}
\item Most of the calibration tests with a single expiry have run into an optimal value different than the original one. It is to be seen if this is a property of the Heston model or if this was due instead to the specific parameter or strike/maturity values being used.


\item It would be interesting to study the properties of the SWIFT implementation proposed for a chosen set of strikes in expression (\ref{step:multiple_strikes2}). We could interpolate the values at all strikes with splines methods that require derivatives, and not only derivative-free ones.

\item Options with very long maturities may hamper the calibration process due to numerical overflows during the pricing step. The problem of long maturities has been tackled with Haar wavelets in \cite{Ortiz-Gracia2013}. It might be worth investigating whether we can do the same with Shannon wavelets.

\item Deep OTM options with very short maturity are challenging to price. The problem seems to be the lack of accuracy of the approximation on the tails of the density function. 
\item Comparison with other calibration methods based on approximation formulae, like for instance the work by \cite{alos15}.
\end{itemize}

\section*{Acknowledgements}
	L. Ortiz-Gracia acknowledges the Spanish Ministry of Economy and Competitiveness for funding under grant PID2019-105986GB-C21.

\bibliography{biblio}{}
\bibliographystyle{plain}

\appendix

\section{Gradient complimentary formulas from \cite{cui17}}\label{app:cui_extra}
Partial derivatives for gradient computation of the Heston model ChF,
\begin{align}
\frac{\partial d}{\partial \rho}=&\frac{\xi \sigma i u}{d}, \\
\frac{\partial A_{2}}{\partial \rho}=&\frac{\sigma i u(2+\xi \tau)}{2 d \nu_{0}}\left(\xi \cosh \frac{d \tau}{2}+d \sinh \frac{d \tau}{2}\right), \\
\frac{\partial B}{\partial \rho}=&\frac{e^{\kappa \tau / 2}}{\nu_{0}}\left(\frac{1}{A_{2}} \frac{\partial d}{\partial \rho}-\frac{d}{A_{2}^{2}} \frac{\partial A_{2}}{\partial \rho}\right), \\
\frac{\partial A_{1}}{\partial \rho}=&\frac{i u\left(u^{2}-i u\right) \tau \xi \sigma}{2 d} \cosh \frac{d \tau}{2}, \\
\frac{\partial A}{\partial \rho}=&\frac{1}{A_{2}} \frac{\partial A_{1}}{\partial \rho}-\frac{A}{A_{2}} \frac{\partial A_{2}}{\partial \rho},\\
\frac{\partial A}{\partial \kappa}=&-\frac{i}{\sigma u} \frac{\partial A}{\partial \rho},\\
\frac{\partial B}{\partial \kappa}=&-\frac{i}{\sigma u}\dfrac{\partial B}{\partial \rho} + \dfrac{B \tau}{2},\\
\frac{\partial d}{\partial \sigma}=& \left( \frac{d}{\sigma} - \dfrac{1}{\xi} \right) \frac{\partial d}{\partial \rho} + \dfrac{\sigma u^2}{d},\\
\frac{\partial A_{1}}{\partial \sigma}=&\frac{\left(u^{2}-i u\right) \tau}{2} \frac{\partial d}{\partial \sigma} \cosh \frac{d \tau}{2}, \\
\frac{\partial A_{2}}{\partial \sigma}=&\frac{\rho}{\sigma} \frac{\partial A_{2}}{\partial \rho}+\frac{2+\tau \xi}{\nu_{0} \tau \xi i u} \frac{\partial A_{1}}{\partial \rho}+\frac{\sigma \tau A_{1}}{2 \nu_{0}}, \\
\frac{\partial A}{\partial \sigma}=&\frac{1}{A_{2}} \frac{\partial A_{1}}{\partial \sigma}-\frac{A}{A_{2}} \frac{\partial A_{2}}{\partial \sigma}.
\end{align}

\section{Strike and maturity test sets}\label{app:strikes}
Set 1 and set 2 are provided in Tables \ref{table:set1} and \ref{table:set2}, respectively. The goal of the set 3 is just to check the behavior of KSWIFT in the worst configuration possible for its speeding up techniques. This scenario is only be applied to KSWIFT, and consists of the same strikes and maturities as set 2. The code implementation of KSWIFT generated for this work receives as inputs a vector of expiry-defined-data (EDD). Each element of the vector of EDD contains a single expiry and a vector of strikes. Thus, set 2 will have all the strikes with the same expiry grouped in a single EDD, and set 3 will have EDD consisting on a single strike. This will enforce full recomputation of the density and payoff coefficients for each strike.
\begin{table}[!h]
\begin{center}
 \begin{tabular}{| c | c | c | c | c |} 
 \hline
 Strike & Strike & Strike & Strike & Strike \\ [0.5ex] 
 \hline
  0.9371 & 0.9956 & 1.0427 & 1.2287 & 1.3939 \\
\hline
  0.8603 & 0.9868 & 1.0463 & 1.2399 & 1.4102 \\
\hline
  0.8112 & 0.9728 & 1.0499 & 1.2485 & 1.4291 \\
\hline
  0.7760 & 0.9588 & 1.0530 & 1.2659 & 1.4456 \\
\hline
  0.7470 & 0.9464 & 1.0562 & 1.2646 & 1.4603 \\
\hline
  0.7216 & 0.9358 & 1.0593 & 1.2715 & 1.4736 \\
\hline
  0.6699 & 0.9175 & 1.0663 & 1.2859 & 1.5005 \\
\hline
  0.6137 & 0.9025 & 1.0766 & 1.3046 & 1.5328 \\
 \hline
\end{tabular}
\caption{Set 1 of strikes and expiries. All the strikes have the same expiry $\tau=0.119047619047619$.}\label{table:set1}
\end{center}
\end{table}

\begin{table}[!h]
\begin{center}
 \begin{tabular}{|c | c | c | c | c | c |} 
 \hline
 Expiry & Strike & Strike & Strike & Strike & Strike \\ [0.5ex] 
 \hline
 0.119047619047619 & 0.9371 & 0.9956 & 1.0427 & 1.2287 & 1.3939 \\
\hline
0.238095238095238 & 0.8603 & 0.9868 & 1.0463 & 1.2399 & 1.4102 \\
\hline
0.357142857142857 & 0.8112 & 0.9728 & 1.0499 & 1.2485 & 1.4291 \\
\hline
0.476190476190476 & 0.7760 & 0.9588 & 1.0530 & 1.2659 & 1.4456 \\
\hline
0.595238095238095 & 0.7470 & 0.9464 & 1.0562 & 1.2646 & 1.4603 \\
\hline
0.714285714285714 & 0.7216 & 0.9358 & 1.0593 & 1.2715 & 1.4736 \\
\hline
1.07142857142857 & 0.6699 & 0.9175 & 1.0663 & 1.2859 & 1.5005 \\
\hline
1.42857142857143 & 0.6137 & 0.9025 & 1.0766 & 1.3046 & 1.5328 \\
 \hline
\end{tabular}
\caption{Set 2 of strikes and expiries.}\label{table:set2}
\end{center}
\end{table}

\end{document}